\documentclass[french,english,aps,preprint,showpacs,showkeys]{revtex4}
\usepackage[T1]{fontenc}
\usepackage[latin9]{inputenc}
\usepackage{xcolor}
\usepackage{amsmath}
\usepackage{amssymb}

\makeatletter
\@ifundefined{textcolor}{}
{%
 \definecolor{BLACK}{gray}{0}
 \definecolor{WHITE}{gray}{1}
 \definecolor{RED}{rgb}{1,0,0}
 \definecolor{GREEN}{rgb}{0,1,0}
 \definecolor{BLUE}{rgb}{0,0,1}
 \definecolor{CYAN}{cmyk}{1,0,0,0}
 \definecolor{MAGENTA}{cmyk}{0,1,0,0}
 \definecolor{YELLOW}{cmyk}{0,0,1,0}
}

\makeatother

\usepackage{babel}
\makeatletter
\addto\extrasfrench{%
   \providecommand{\fg}{\ifdim\lastskip>\z@\unskip\fi~\frqq}%
}

\makeatother
\begin{document}

\title{Exact solutions of a two-dimensional Duffin-Kemmer-Petiau oscillator
subject to a coulomb potential in the gravitational field of cosmic
string}

\author{Abdelmalek Boumali}
\email{boumali.abdelmalek@gmail.com}

\affiliation{Laboratoire de Physique Appliquée et Théorique, \\
University Larbi Tébessi -Tébessa-, 12000, W. Tébessa, Algeria.}

\author{Nadjette Messai}
\email{nadjette.messai@gmail.com}

\affiliation{Laboratoire de Physique Appliquée et Théorique, \\
University Larbi Tébessi -Tébessa-, 12000, W. Tébessa, Algeria.}
\begin{abstract}
In this paper, the problem of a two-dimensional Duffin-Petiau-Kemmer
(DKP) oscillator in the presence of a coulomb potential in the cosmic
string background is solved. The eigensolutions of the problem in
question have been found, and the influence of the Coulomb potential
in the presence of the gravitational field of cosmic string has been
analyzed.
\end{abstract}


\maketitle

\section{introduction}

The analysis of gravitational interactions with a quantum mechanical
system has recently attracted attention in particle physics and has
been an active field of research.The general way to understand the
interaction between relativistic quantum mechanical particles and
gravity is to solve the general relativistic form of their wave equations.
These solutions are valuable tools for examining and improving models
and numerical methods for solving complicated physical problems.

In the conventional relativistic approach, the interaction of $S=0$
and $S=1$ hadrons with different nuclei has been described by the
second-order Klein-Gordon (KG) equation for S=0 and Proca equation
for S=1 particles. It is well known that is very difficult to tackle
these second-order equations mathematically and to derive the physics
behind them. Therefore, considerable interest in recent years has
been devoted to examining the interactions of $S=0$ and $S=1$ hadrons
with nuclei by using the first-order relativistic Duffin-Kemmer-Petiau
(DKP) equation \cite{1}.

One important question related to DKP equation concerns the equivalence
between its spin $0$ and $1$ sectors and the theories based on the
second-order KG and Proca equations\cite{2}. Historically, the loss
of interest in the DKP stems from the equivalence of the DKP approach
to the Klein-Gordon (KG) and Proca descriptions in on-shell situations,
in addition to the greater algebraic complexity of the DKP formulation.
However, in the 1970s, this supposed equivalence was question in several
situations involving breaking of symmetries and hadronic possess,
showing that in some cases, the DKP and KG theories can give different
results. Moreover, the DKP equation appears to be richer than the
KG equation if the interactions are introduced. In this context, alternative
DKP-based models were proposed for the study of meson-nucleus interactions,
yielding a better adjustment to the experimental data when compared
to the KG-based theory\cite{3}. In the same direction, approximation
techniques formerly developed in the context of nucleon-nucleus scattering
were generalized, giving a good description for experimental data
of meson-nucleus scattering\cite{4}. The deuteron-nucleus scattering
was also studied using DKP equation, motivated by the fact that this
theory suggest a spin$1$ structure from combining two spin-$\frac{1}{2}$\cite{5}.
In addition, we can cite the works of \cite{6,7} on the meson-nuclear
interaction and the relativistic model of $\alpha-$nucleus elastic
scattering where they have been treated by the formalism of the DKP
theory. Recently, there is a renewed interest in the DKP equation.
It has been studied in the context of quantum chromodynamics (QCD)
\cite{8}, covariant Hamiltonian formalism \cite{9}, in the causal
approach\cite{10,11}, in the context of five-dimensional Galilean
invariance\cite{12}, in the scattering of $K^{+}$ nucleus\cite{13},
in the presence of the Aharonov-Bohm potential\cite{14,15}, in the
Dirac oscillator interaction\cite{16}, in the study of thermodynamics
properties\cite{17}, on the supersymmetric\cite{18}, and finally
in the presence of some shape of interactions\cite{19,20,21,22,23,24,25,26,27,28,29,30,31}.
Theses examples in some case break the equivalence between the theories
based on the DKP equation and KG and Proca equations.

The Dirac oscillator was for the first time studied by Itô and Carriere
\cite{32}. On the other side, Moshinsky and Szczepaniak were the
first who introduced an interesting term in the Dirac equation. More
specifically, they suggested to substitute in the free Dirac equation
the momentum operator $\vec{p}$ like $\vec{p}-im\omega\beta\vec{r}$.
They could obtain a system in which the positive energy states have
a spectrum similar to the one of the non-relativistic harmonic oscillator
\cite{33}. Recently, this interaction has particularly got more interest.
It is reviewed, because of the interest in the many different domain
in physics (see Ref. \cite{33} and references therein).

The topological defects plays an important role in physical properties
of systems, and they appear in gravitation as monopoles, strings and
walls \cite{34,35,36,37,38,39,40,41,42,43}. Among them, cosmic strings
and monopoles seem to be the best candidates to be observed. The former
are linear defects, and the space-time produced by an idealized cosmic
string is locally flat, however, globally conical, with a planar angle
deficit determined by the string tension.

The well-known procedure to introduce the coupling between a charged
particle and electromagnetic fields in the DKP equation, is through
the minimal coupling. Dosch, Jensen and Müller in 1971 proposed another
procedure by making a modification in the mass term in the form: $m\rightarrow m+S\left(\vec{r}\right)$
where $S\left(\vec{r}\right)$ is the scalar potential \cite{45}.
\textcolor{black}{This new formalism has been used by Soff et al \cite{46}
to analyze the Dirac equation in the presence of a Coulomb potential
and a static scalar potential. Bergerhoff and Soff \cite{47} show
that, in contrast to the minimal coupling of the electromagnetic potentials
where it is correlated with the momentum, a scalar potentials, which
are an invariant Lorentz's scalar, are coupled to the mass of a particle
in the Dirac equation and thus act effectively as a position dependent
mass. They have shown that, when a scalar external is coupled instead
of a vector Coulomb potential, the no present of Klein\textquoteright s
paradox and consequently the spontaneous pair creation . This implies,
that for an arbitrary scalar potential one can always find the bound
states in the gap between $+m_{e}c^{2}$ and $-m_{e}c^{2}$, which
is not the case for the usual Coulomb potential, which is coupled
to the Dirac field by replacing $\vec{p}$ by $\vec{p}-(e/c)\vec{A}$
in the Dirac equation.}\textcolor{red}{\emph{ }}More recently, Medeiros
et al \cite{44} have been used this formalism to study Relativistic
quantum dynamics of a charged particle in cosmic string space-time
in the presence of magnetic field and scalar potential. In the same
context, Bakke and his co-workers \cite{40} have been studied several
problems by using this new approach. They showed that this modification
in the mass term gives rise to a position-dependent mass for a relativistic
particle ( see Ref.\cite{38} and references therein). Following Bakke
\cite{40}, this method has been used in different situation such
as : (i) the quark\textendash antiquark interaction, (ii) analysis
of the behavior of a Dirac particle in both static scalar and Coulomb
potentials, (iii) in a relativistic scalar particle in the cosmic
string space time, and (iv) finally in the Klein-Gordon oscillator
subject to a Coulomb potential. Following Medeiros et al \cite{44},
if one wants to investigate the relativistic quantum motion of a charged
particle in the presence of electromagnetic and scalar potentials,
both procedures, the minimal coupling and a modification in the mass
term, should be taken into account. The problem of the wave functions
of particles subject to different confining potentials as a Coulomb
potential, whose exact solution as well established, has been made
by replacing $\vec{p}$ by $\vec{p}-\frac{e}{c}\vec{A}$ in the relativistic
particle equation's: this potential is a time-like component of the
electromagnetic vector potential. 

The principal aim of this paper is to solve the DKP oscillator in
a background produced by topological defects, such as cosmic strings
in the presence of a Coulomb potential. The introduction of the Coulomb
potential in the DKP equation will be made by using the minimal coupling
procedure. The structure of this article is as follows: In Sect. II,
we briefly review the DKP equation in cosmic string background subject
to the Coulomb potential. In Sect III, the eigensolutions have been
obtained for both massive Spin-0 and spin-1 particles. Sect IV present
our conclusion.

\section{The DKP oscillator In Cosmic String Background}

In this section, we discuss the DKP oscillator in cosmic string space-time
described by the metric \cite{48,49}
\begin{equation}
ds^{2}=-dt^{2}+d\rho^{2}+\alpha^{2}\rho^{2}d\phi^{2}+dz^{2},\label{eq:1}
\end{equation}
with $-\infty<\left(t,z\right)<+\infty$, $0<\rho\leq\infty$ and
$0\leq\phi\leq2\pi$. The parameter $\alpha$ is the deficit angle
associated with conical geometry obeying $\alpha=1-4\eta$, and $\eta$
is the linear mass density of the string in natural unite $\hbar=c=1$. 

The DKP equation in curved space-time is given by \cite{50,51,52,53}
\begin{equation}
\left[i\tilde{\beta^{\mu}}\left(\partial_{\mu}+\frac{1}{2}\omega_{\mu ab}S^{ab}\right)-m\right]\psi=0,\label{eq:2}
\end{equation}
where $\tilde{\beta^{\mu}}$ are the DKP matrices in curved space,
and they satisfy the following relations:
\begin{equation}
\beta^{a}\beta^{b}\beta^{c}+\beta^{c}\beta^{b}\beta^{a}=\beta^{a}\eta^{bc}+\beta^{c}\eta^{ba}.\label{eq:3}
\end{equation}
These matrices are related to the flat space-time $\beta^{a}$ as:
$\tilde{\beta^{\mu}}=e_{a}^{\mu}\beta^{a}$ with the following tetrad
relations: The $\beta^{a}$ are $5\times5$ matrices in the spin-0
representation and $10\times10$ matrices in the spin-1 representation.
In our case, the $\beta^{a}$ matrices are chosen as follows:\cite{54,55}
\begin{itemize}
\item for the spin-0 representation 
\begin{equation}
\begin{array}{cc}
\beta^{0}=\begin{pmatrix}\theta_{2\times2} & 0_{3\times3}\\
0_{3\times2} & 0_{2\times3}
\end{pmatrix}, & \beta^{i}=\begin{pmatrix}0_{2\times2} & \rho_{2\times3}^{i}\\
-\rho_{3\times2}^{iT} & 0_{3\times3}
\end{pmatrix},\end{array}\left(i=1,2\right),\label{eq:5}
\end{equation}
\end{itemize}
with 
\begin{equation}
\theta_{2\times2}=\begin{pmatrix}0 & 1\\
1 & 0
\end{pmatrix},\,\rho^{1}=\begin{pmatrix}-1 & 0 & 0\\
0 & 0 & 0
\end{pmatrix},\,\rho^{2}=\begin{pmatrix}0 & -1 & 0\\
0 & 0 & 0
\end{pmatrix}.\label{eq:6}
\end{equation}

\begin{itemize}
\item for Spin-1 representation 
\begin{equation}
\beta^{0}=\begin{bmatrix}0_{3\times3} & 0_{3\times3} & -I_{3\times3} & 0_{3\times1}^{+}\\
0_{3\times3} & 0_{3\times3} & 0_{3\times3} & 0_{3\times1}^{+}\\
-I_{3\times3} & 0_{3\times3} & 0_{3\times3} & 0_{3\times1}^{+}\\
0_{1\times3} & 0_{1\times3} & 0_{1\times3} & 0
\end{bmatrix},\beta^{k}=\begin{bmatrix}0_{3\times3} & 0_{3\times3} & -I_{3\times3} & iK^{k+}\\
0_{3\times3} & 0_{3\times3} & S_{3\times3}^{k} & 0_{3\times1}^{+}\\
-I_{3\times3} & -S_{3\times3}^{k} & 0_{3\times3} & 0_{3\times1}^{+}\\
iK^{k} & 0_{1\times3} & 0_{1\times3} & 0
\end{bmatrix}\,\left(k=1,2\right),\label{eq:7}
\end{equation}
\end{itemize}
where 
\[
\begin{array}{cc}
0_{3\times3}=\begin{pmatrix}0 & 0 & 0\\
0 & 0 & 0\\
0 & 0 & 0
\end{pmatrix}, & I=\begin{pmatrix}1 & 0 & 0\\
0 & 1 & 0\\
0 & 0 & 1
\end{pmatrix},\end{array}
\]
and
\begin{equation}
S^{1}=i\begin{pmatrix}0 & 0 & 0\\
0 & 0 & -1\\
0 & 1 & 0
\end{pmatrix},\,S^{2}=i\begin{pmatrix}0 & 0 & 1\\
0 & 0 & 0\\
-1 & 0 & 0
\end{pmatrix},\label{eq:8}
\end{equation}
with
\[
,K^{1}=\begin{pmatrix}1 & 0 & 0\end{pmatrix},\,K^{2}=\begin{pmatrix}0 & 1 & 0\end{pmatrix},
\]
and $S^{ab}=\left[\beta^{a},\beta^{b}\right]$. The spin connection
$\omega_{\mu ab}$ obeys the relations
\begin{equation}
\begin{array}{cc}
\omega_{\mu ab}=e_{al}e_{b}^{j}\Gamma_{j\mu}^{l}-e_{b}^{j}\partial_{\mu}e_{aj}, & \omega_{\mu ab}=-\omega_{\mu ba},\end{array}\label{eq:10}
\end{equation}
with
\begin{equation}
\Gamma_{j\mu}^{l}=\frac{1}{2}g^{l\lambda}\left(\partial_{l}g_{\mu\lambda}+\partial_{\mu}g_{\lambda j}-\partial_{\lambda}g_{\mu j}\right),\label{eq:11}
\end{equation}
are the Christoffel symbols or the affine connections. 

The introduction of the interaction will do by modifying the term
$p_{\mu}\rightarrow p_{\mu}-qA_{\mu}$ where $q$ is the electric
charge, and $A_{\mu}=\left(-A_{0},\overrightarrow{A}\right)$ is the
electromagnetic 4-vector potential. Takes into account the substitution
$\partial_{\rho}\rightarrow\partial_{\rho}+m\omega\rho$ into Eq.
(\ref{eq:2}), and the following form of the Coulomb potential \cite{56,57,58}
\begin{equation}
qA^{0}=\frac{f}{\rho}=\pm\frac{\left|f\right|}{\rho},\label{eq:12}
\end{equation}
where $f$ is a constant, Eq.(\ref{eq:2}) can be written as
\begin{equation}
\left\{ i\tilde{\beta^{0}}\left(\partial_{0}-iqA_{0}\right)+i\tilde{\beta^{1}}\left(\partial_{\rho}+m\omega\rho\varsigma\right)+i\tilde{\beta^{2}}\left(\partial_{\phi}+\alpha\left[\beta^{1},\beta^{2}\right]\right)-m\right\} \psi=0,\label{eq:13}
\end{equation}
with $\varsigma=2\left(\beta^{0}\right)^{2}-I$ and $\varsigma^{2}=I.$

In what follow, this equation will be used to extract the eigensolutions
of a both massive spin-0 and spin-1 particles.

\section{The eigensolutions of a two-dimensional DKP oscillator in cosmic
string background}

\subsection{\textit{case of spin zero}}

The two dimensional DKP equation in cosmic string space-time subject
to the Coulomb potential is given by Eq. (\ref{eq:13}), where $m_{0}$
is the mass of particles of spin-0. The stationary state $\psi$ is
a five-component wave function of the DKP equation, which can be written
as 
\begin{equation}
\psi=\begin{pmatrix}\psi_{1}, & \psi_{2}, & \psi_{3}, & \psi_{4}, & \psi_{5}\end{pmatrix}^{T}\label{eq:14}
\end{equation}
Substituting (\ref{eq:5}), (\ref{eq:6}) and (\ref{eq:14}) into
(\ref{eq:13}), we obtain 
\begin{equation}
\left(E+\frac{f}{\rho}\right)\psi_{2}-i\left(\partial_{\rho}-m_{0}\omega\rho+\frac{1}{\rho}\right)\psi_{3}-i\frac{\partial_{\phi}}{\alpha\rho}\psi_{4}-m_{0}\psi_{1}=0,\label{eq:15}
\end{equation}
\begin{equation}
\left(E+\frac{f}{\rho}\right)\psi_{1}-m_{0}\psi_{2}=0,\label{eq:16}
\end{equation}
\begin{equation}
i\left(\partial_{\rho}+m\omega\rho\right)\psi_{1}-m_{0}\psi_{3}=0,\label{eq:17}
\end{equation}
\begin{equation}
i\frac{\partial_{\phi}}{\alpha\rho}\psi_{1}-m_{0}\psi_{4}=0,\label{eq:18}
\end{equation}
\begin{equation}
-m_{0}\psi_{5}=0.\label{eq:19}
\end{equation}
From these equations, we get the following relations
\begin{equation}
\psi_{2}=\frac{\left(E+\frac{f}{\rho}\right)}{m_{0}}\psi_{1},\label{eq:20}
\end{equation}
\begin{equation}
\psi_{3}=i\frac{\left(\partial_{\rho}+m_{0}\omega\rho\right)}{m_{0}}\psi_{1},\label{eq:21}
\end{equation}
\begin{equation}
\psi_{4}=\frac{i}{m_{0}\alpha\rho}\partial_{\phi},\label{eq:22}
\end{equation}
\begin{equation}
\psi_{5}=0.\label{eq:23}
\end{equation}
Putting Eqs. (\ref{eq:20}), (\ref{eq:21}) and (\ref{eq:22}) into
Eq. (\ref{eq:15}), we have
\begin{equation}
\left\{ \left(E+\frac{f}{\rho}\right)^{2}+\left(\partial_{\rho}-m\omega\rho+\frac{1}{\rho}\right)\left(\partial_{\rho}+m\omega\rho\right)+\frac{1}{\alpha^{2}\rho^{2}}\partial_{\phi}^{2}-m_{0}^{2}\right\} \psi_{1}=0.\label{eq:24}
\end{equation}
Choosing as Ansatz $\psi_{1}=e^{iJ\phi}\chi\left(\rho\right)$, and
after simple algebraic manipulations, we arrive at 
\begin{equation}
\left[\frac{\partial^{2}}{\partial\rho^{2}}+\frac{1}{\rho}\frac{\partial}{\partial\rho}-\frac{\beta^{2}}{\rho^{2}}+\frac{2Ef}{\rho}-m_{0}^{2}\omega^{2}\rho^{2}+\nu\right]\chi\left(\rho\right)=0,\label{eq:25}
\end{equation}
with 
\begin{equation}
\begin{array}{ccc}
\nu=E^{2}-m_{0}^{2}+2m_{0}\omega, & \beta^{2}=\lambda^{2}-f^{2},\lambda= & \frac{J}{\alpha}\end{array}.\label{eq:26}
\end{equation}
Now let us make a change of variable $\zeta=\sqrt{m_{0}\omega}\rho$:
in this case equation (\ref{eq:25}) becomes
\begin{equation}
\left[\frac{\partial^{2}}{\partial\xi^{2}}+\frac{1}{\xi}\frac{\partial}{\partial\xi}-\frac{\beta^{2}}{\xi^{2}}+\frac{\delta}{\xi}-\xi^{2}+\frac{\nu}{m_{0}\omega}\right]\chi\left(\xi\right)=0.\label{eq:27}
\end{equation}
Taking the following substitutions \cite{56,57,58,59}
\begin{equation}
\delta=\frac{2Ef}{\sqrt{m_{0}\omega}},\,\chi\left(\xi\right)=e^{-\frac{\xi^{2}}{2}}\xi^{\begin{vmatrix}\varsigma\end{vmatrix}}H\left(\xi\right),\label{eq:27-1}
\end{equation}
where we have write $\chi\left(\xi\right)$ as unknown function $H\left(\xi\right)$,
Eq. (\ref{eq:27}) can be rewritten by
\begin{equation}
H^{''}\left(\xi\right)+\left\{ \left(2\begin{vmatrix}\beta\end{vmatrix}+1\right)\frac{1}{\xi}-2\xi\right\} H^{'}\left(\xi\right)+\left\{ \frac{\nu}{m\omega}-2\begin{vmatrix}\beta\end{vmatrix}-2+\frac{\delta}{\xi}\right\} H\left(\xi\right)=0.\label{eq:29}
\end{equation}
The last equation is a biconfluent Heun function\cite{57,58,59} 
\begin{equation}
H\left(\xi\right)=H\left(2\begin{vmatrix}\beta\end{vmatrix},0,\frac{\nu}{m\omega},2\delta,-\xi\right).\label{eq:30}
\end{equation}
In order to solve Eq. (\ref{eq:29}), we use the Frobenius method
\cite{60,61,62,63}: Eq.(\ref{eq:30}) can be written as a power series
expansion around the origin as
\begin{equation}
H\left(\xi\right)={\displaystyle \sum_{j=0}^{\infty}}a_{j}\xi^{j}.\label{eq:31}
\end{equation}
Substituting the series (\ref{eq:31}) into equation (\ref{eq:29}
), we obtain the following recurrence relation:
\begin{equation}
a_{j+2}=-\frac{\delta}{\left(j+2\right)\left(j+1+\gamma\right)}a_{j+1}-\frac{\theta-2j}{\left(j+2\right)\left(j+1+\gamma\right)}a_{j},\label{eq:32}
\end{equation}
where $\gamma=2\begin{vmatrix}\beta\end{vmatrix}+1$ and $\theta=\frac{\nu}{m\omega}-2\begin{vmatrix}\beta\end{vmatrix}-2$.
By starting with $a_{0}=1$ and using the relation (\ref{eq:32}),
we can calculate the other coefficients of the power series expansion
as follows
\begin{align}
a_{1} & =-\frac{\delta}{\gamma}=-\frac{2Ef}{\sqrt{m_{0}\omega}}\frac{1}{2\begin{vmatrix}\beta\end{vmatrix}+1},\label{eq:33}
\end{align}
\begin{align}
a_{2} & =\frac{\delta^{2}}{2\gamma\left(1+\gamma\right)}-\frac{\theta}{2\left(1+\gamma\right)}=\frac{2E^{2}f^{2}}{m_{0}\omega}\frac{1}{\left(2\begin{vmatrix}\beta\end{vmatrix}+1\right)\left(2\begin{vmatrix}\beta\end{vmatrix}+2\right)}-\frac{\theta}{2\left(2\begin{vmatrix}\beta\end{vmatrix}+2\right)}.\label{eq:34-1}
\end{align}
Thus, the bound state solutions can be obtained by imposing the conditions
where power series becomes a polynomial of degree $n$. This happens
when:
\begin{equation}
\theta=2n,\,a_{n+1}=0.\label{eq:34}
\end{equation}
with now $n=1,2,3,\ldots$. From Eq. (\ref{eq:34}), the eigenvalues
are given by
\begin{equation}
E_{n,J}^{2}=m_{0}^{2}+2m_{0}\omega_{n,J}\left(n+\begin{vmatrix}\sqrt{\frac{J^{2}}{\alpha^{2}}-f^{2}}\end{vmatrix}\right).\label{eq:36}
\end{equation}
Following Eq. (\ref{eq:36}), the spectrum of energy of the DKP oscillator
subject to the Coulomb potential in the presence of the gravitational
shows two interesting results: (i) firstly, all levels of energy are
not degenerate, and (ii) secondly, the introduction of the Coulomb
potential in the DKP oscillator modified the relativistic energy levels.
This influence yields the ground state of DKP oscillator to be defined
by the quantum number $n=1$ in contrast to the quantum number $n=0$.
This situation is similar to the case of the Klein-Gordon oscillator
studied by Bakke et al \cite{36,37,38,39}. So we can extended their
physical interpretation in our case. Thus, when the condition $a_{n+1}=0$
imposed (Eq. (\ref{eq:34})), we obtain a polynomial of degree $n$
the power series expansion given in Eq. (\ref{eq:31}). So, we can
assume that frequency $\omega_{n,J}$ of the DKP oscillator can be
adjusted in order to satisfied the condition $a_{n+1}=0$. As a consequence,
the quantum number of the system $n$ restrict the possible values
of the angular frequency. According this, there are values of the
angular frequency which are not allowed in the system. Now, let us
examine the case of the ground state $n=1$: the condition $a_{n+1}=0$
yields $a_{2}=0$. So, by using Eq. (\ref{eq:32}), frequency $\omega_{n,J}$
is given by
\begin{equation}
\omega_{1,J}=\frac{2f^{2}E_{1,J}^{2}}{m_{0}\left(2\left|\beta\right|+1\right)},\label{eq:36-1}
\end{equation}
which corresponds to the possible values of the angular frequency
of the DKP oscillator in the ground state. The energy levels corresponding
of this ground state are written by
\begin{equation}
E_{n,J}^{2}=\pm\frac{m_{0}}{\sqrt{1-\frac{4f^{2}}{2\left|\beta\right|+1}\left(1+\left|\sqrt{\frac{J^{2}}{\alpha^{2}}-f^{2}}\right|\right)}}.\label{eq: 36.2}
\end{equation}
Therefore, the effects of the Coulomb potential on the spectrum of
energy of the DKP oscillator in the presence of a cosmic string is
given by a change of the energy levels, where the ground state is
defined by the quantum number $n=1$. Moreover, the values of the
angular frequency of the DKP oscillator are restricted to a set of
values in which allow us to obtain a polynomial solution to the biconfluent
Heun series \cite{36,37,38,39}. 

Finally, when we take the limit $f\rightarrow0$ (i.e., vanishing
of the Coulomb potential which here is chosen as time-like component
of $A^{\mu}$), we recover the exact result of scalar bosons in a
cosmic string background \cite{42}. Now, if we take a both limits
$\alpha\rightarrow1$ with $f\rightarrow0$ , we obtain the same result
found in the case of a two-dimensional DKP oscillator in Minkowski
space-time\cite{54}.

\subsection{\textit{case of spin one}}

The two dimensional DKP equation in cosmic string space-time subject
to a Coulomb potential is
\begin{equation}
\left\{ i\tilde{\beta^{0}}\left(\partial_{0}-iqA_{0}\right)+i\tilde{\beta^{1}}\left(\partial_{\rho}+m\omega\rho\varsigma\right)+i\tilde{\beta^{2}}\left(\partial_{\phi}+\alpha\left[\beta^{1},\beta^{2}\right]\right)-M\right\} \psi=0,\label{eq:37}
\end{equation}
where $M$ is the mass of spin-1 particles. The stationary state $\psi$
is a ten-component wave function of the DKP equation with
\begin{equation}
\psi=\begin{pmatrix}\psi_{1}, & \psi_{2}, & \psi_{3}, & \psi_{4}, & \psi_{5}, & \psi_{6}, & \psi_{7}, & \psi_{8}, & \psi_{9}, & \psi_{10}\end{pmatrix}^{T}.\label{eq:38}
\end{equation}
Substituting Eqs. (\ref{eq:7}), (\ref{eq:8}) and (\ref{eq:38})
into Eq. (\ref{eq:37}), we found the following system of equations
\begin{equation}
-\left(E+\frac{f}{\rho}\right)\psi_{7}+\left(\partial_{\rho}-M\omega\rho\right)\psi_{10}-M\psi_{1}=0,\label{eq:39}
\end{equation}
\begin{equation}
-\left(E+\frac{f}{\rho}\right)\psi_{8}+\frac{\partial_{\phi}}{\alpha\rho}\psi_{10}-M\psi_{2}=0,\label{eq:40}
\end{equation}
\begin{equation}
-\left(E+\frac{f}{\rho}\right)\psi_{9}-M\psi_{3}=0,\label{eq:41}
\end{equation}
\begin{equation}
-\frac{\partial_{\phi}}{\alpha\rho}\psi_{9}-M\psi_{4}=0,\label{eq:42}
\end{equation}
\begin{equation}
\left(\partial_{\rho}+M\omega\rho\right)\psi_{9}-M\psi_{5}=0,\label{eq:43}
\end{equation}
\begin{equation}
\frac{\partial_{\phi}}{\alpha\rho}\psi_{7}-\left(\partial_{\rho}+M\omega\rho+\frac{1}{\rho}\right)\psi_{8}-M\psi_{6}=0,\label{eq:44}
\end{equation}
\begin{equation}
-\left(E+\frac{f}{\rho}\right)\psi_{1}+\frac{\partial_{\phi}}{\alpha\rho}\psi_{6}-M\psi_{7}=0,\label{eq:45}
\end{equation}
\begin{equation}
-\left(E+\frac{f}{\rho}\right)\psi_{2}-\left(\partial_{\rho}-M\omega\rho\right)\psi_{5}-M\psi_{8}=0,\label{eq:46}
\end{equation}
\begin{equation}
-\left(E+\frac{f}{\rho}\right)\psi_{3}+\left(\partial_{\rho}-M\omega\rho+\frac{1}{\rho}\right)\psi_{5}-\frac{\partial_{\phi}}{\alpha\rho}\psi_{4}-M\psi_{9}=0,\label{eq:47}
\end{equation}
\begin{equation}
-\left(\partial_{\rho}+M\omega\rho-\frac{1}{\rho}\right)\psi_{1}-\frac{\partial_{\phi}}{\alpha\rho}\psi_{2}-M\psi_{10}=0.\label{eq:48}
\end{equation}
From Eqs. (\ref{eq:41}), (\ref{eq:42}) and (\ref{eq:43}), we have
\begin{equation}
\psi_{3}=-\frac{\left(E+\frac{f}{\rho}\right)}{M}\psi_{9},\label{eq:49}
\end{equation}
\begin{equation}
\psi_{4}=-\frac{1}{M}\frac{\partial_{\phi}}{\alpha\rho}\psi_{9},\label{eq:50}
\end{equation}
\begin{equation}
\psi_{5}=\frac{1}{M}\left(\partial_{\rho}+M\omega\rho\right)\psi_{9}.\label{eq:51}
\end{equation}
Putting these equations into Eq. (\ref{eq:47}), we arrive at the
following equation for $\psi_{9}$
\begin{equation}
\left\{ \left(E+\frac{f}{\rho}\right)^{2}+\left(\partial_{\rho}-M\omega\rho+\frac{1}{\rho}\right)\left(\partial_{\rho}+M\omega\rho\right)+\frac{\partial_{\phi}^{2}}{\alpha^{2}\rho^{2}}-M^{2}\right\} \psi_{9}=0\label{eq:52}
\end{equation}
For the other components, it very difficult to decouple the system
of equations above as was done for $\psi_{9}$. In order to overcome
this problem, we use the same method as in Ref.\cite{54}: if we choose

\begin{equation}
\psi_{1}=\psi_{2}=0,\label{eq:54}
\end{equation}
we obtain
\begin{equation}
\psi_{6}=\psi_{7}=\psi_{8}=\psi_{10}=0.\label{eq:55}
\end{equation}
Now, Considering the following Ansatz for the component $\psi_{9}=e^{iJ\phi}\varphi\left(\rho\right)$,
we find
\begin{equation}
\left[\frac{\partial^{2}}{\partial\rho^{2}}+\frac{1}{\rho}\frac{\partial}{\partial\rho}-\frac{\beta^{2}}{\rho^{2}}+\frac{2Ef}{\rho}-M^{2}\omega^{2}\rho^{2}+\nu\right]\varphi\left(\rho\right)=0,\label{eq:53}
\end{equation}
with 
\[
\begin{array}{ccc}
\nu=E^{2}-M^{2}+2M\omega, & \beta^{2}=\lambda^{2}-f^{2},\lambda= & \frac{J}{\alpha}.\end{array}
\]
Eq (\ref{eq:53}) is similar to the Eq. (\ref{eq:25}) for the case
of spin-0. Consequently, the eigensolutions are
\begin{equation}
E_{n}^{2}=M^{2}+2M\omega_{n}\left(n+\begin{vmatrix}\sqrt{\frac{J^{2}}{\alpha^{2}}-f^{2}}\end{vmatrix}\right),\label{eq:56}
\end{equation}
\begin{equation}
\psi_{9}\left(\xi^{'}\right)=e^{-\frac{\xi^{'2}}{2}}\xi^{'\begin{vmatrix}\varsigma\end{vmatrix}}H\left(2\begin{vmatrix}\beta\end{vmatrix},0,\frac{\nu}{m\omega},2\delta^{'},-\xi^{'}\right),\label{eq:57}
\end{equation}
with $\zeta^{'}=\sqrt{M\omega}\rho$ and $\delta^{'}=\frac{2Ef}{\sqrt{M\omega}}$. 

As in the case of spin-0 particles, and from Eq. (\ref{eq:56}), the
following remarks can be made: (i) the eigenvalues of particles of
spin-1 have the same form that for the case of particles of spin-0,
and (ii) all energy levels are not degenerate due the presence of
the gravitational field of cosmic string, and finally (iii) the introduction
of a Coulomb potential in DKP oscillator modified the relativistic
energy levels. This influence yields the ground state of DKP oscillator
to be defined by the quantum number n=1 in contrast to the quantum
number $n=0$. In the limit where $\alpha\rightarrow1$ and $f\rightarrow0$,
we find the same result as in the case of a two-dimensional DKP oscillator
in Minkowski space-time \cite{54}.

\section{Conclusion}

In this work, we have investigated the influence of the topological
defects due to the cosmic strings space-time on the DKP oscillator
subject to the scalar potential such as a Coulomb potential which
is a time-like of the electromagnetic vector potential. The eigenvalues
and eigenfunctions depend explicitly on the non local parameter of
the space-time under consideration even thought it is locally flat.
Contrarily in the case of DKP oscillator for the flat space, the presence
of the topological defects breaks the degeneracy of the spectrum of
the DKP oscillator. In addition, we have seen that the presence of
the Coulomb potential modify the spectrum of energy of the DKP oscillator.
In both cases, the ground state of the system is determined by the
quantum number $n=1$ instead of the quantum number $n=0$. As consequently,
the values of the angular frequency of the DKP oscillator, in both
cases, are restricted to a set of values in which allow us to obtain
a polynomial solution to the biconfluent Heun series.

\end{document}